\documentclass[pre,twocolumn,showpace,superscriptaddress]{revtex4}
\usepackage{graphicx}
\begin{document}
\title{Hole--defect chaos in the one--dimensional complex
  Ginzburg--Landau equation}

\author{Martin Howard}
\affiliation{Department of Mathematics, Imperial
College London, South Kensington Campus, London SW7
2AZ, United Kingdom}

\author{Martin van Hecke}
\affiliation{Kamerlingh Onnes Lab,
Universiteit Leiden, PO Box 9504, 2300 RA Leiden, The
Netherlands}

\date{\today}
\begin{abstract}
  We study the spatiotemporally chaotic dynamics of holes and defects in
  the 1D complex Ginzburg--Landau equation (CGLE). 
  We focus particularly on the self--disordering dynamics of
  holes and on the variation in defect profiles. 
  By enforcing identical defect
  profiles and/or smooth plane wave backgrounds, we are able to 
  sensitively probe the causes of the spatiotemporal chaos.
  We show that the coupling of the holes to a self--disordered
  background is the dominant mechanism. We analyze a lattice model for
  the 1D CGLE, incorporating this self--disordering. Despite its
  simplicity, we show that the model retains the essential
  spatiotemporally chaotic behavior of the full CGLE.
\end{abstract}


\maketitle

\section{Introduction}

The formation of local structures and the occurrence of
spatiotemporal chaos are among the most striking features of pattern
forming systems. The  complex Ginzburg--Landau equation (CGLE)
\begin{equation}
  \partial_t A = A + (1+ ic_1) \partial_x^2 A - (1-i c_3) |A|^2 A \label{cgle}
\end{equation}
provides a particularly rich example of these phenomena. The CGLE is
the amplitude equation describing pattern formation near a Hopf
bifurcation \cite{CH,AK}, and exhibits an extremely wide range of 
behaviors as a function of $c_1$ and $c_3$
\cite{CH,AK,chao1,phasediagram,mvh,maw,mvhmh}.

\begin{figure}
\includegraphics[width=8.5cm]{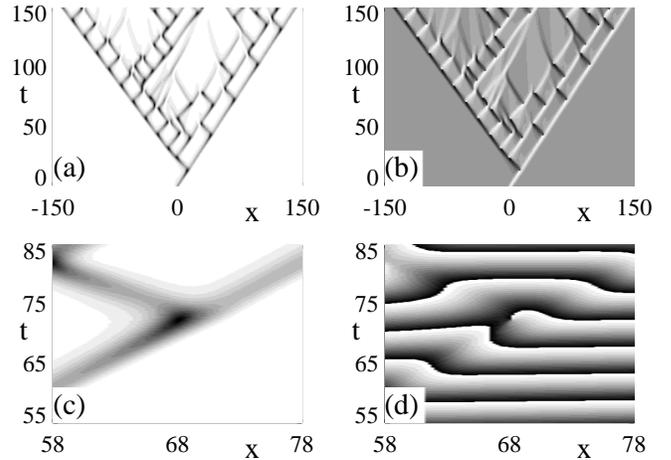}
\caption{Illustration of the main phenomenology of hole--defect chaos
(after Refs.~\cite{phasediagram,mvh,mvhmh}). (a--b) Space--time grey--scale
plots showing the invasion of a plane wave state by hole--defect
chaos: (a) $|A|$ (dark: $A\!\approx\!0$) and (b) $q:=\partial_x \psi$
(light: $q\!>\!0$, grey: $q\!\approx\!0$, dark: $q\!<\!0$). The
propagating objects are incoherent holes, which dynamically connect
the defects (the black dots in (a)). Parameter values are
$c_1\!=\!0.6, c_3\!=\!1.4$ , with an initial condition given by
Eq.~(\ref{peakeq}), with $\gamma\!=\!1, q_{ex}\!=\!-0.03$.  The
non--zero $q_{ex}$ breaks the left--right symmetry and results in the
differing periods of the left and right moving edge holes. (c)
Close--up of $|A|$ and (d) close--up of the complex phase $\psi$,
showing in detail how a hole generates a phase defect that in turn
generates two daughter holes. }\label{fig2}
\end{figure}

{\em Defects} and {\em holes} are local structures that play a crucial
role in the intermediate regime between laminar states (small $c_1$,
$c_3$) and hard chaos (large $c_1$, $c_3$). Isolated defects occur
when $A$ goes through zero, where the complex phase $\psi:= \arg(A)$
is no longer defined. Homoclinic holes are localized propagating
``phase twists'', which are linearly unstable. As illustrated in
Fig.~\ref{fig2}, holes and defects are intimately connected: defects
can give rise to ``holes'', which may then evolve to generate defects,
from which further holes can be born, sometimes generating
self--sustaining patterns. For more details see
Refs.~\cite{mvh,mvhmh}.

The aim of our paper is to understand and model the spatiotemporally
chaotic hole--defect behavior of the 1D CGLE, built on the local
interactions and dynamics of the holes and defects. Given the strength
of the initial phase twist that generates a hole, and the wavenumber of
the state into which it propagates, the hole lifetime $\tau$ turns out
to be the key feature. Surprisingly, the initial phase twist and
invaded state play very different roles. For hole--defect chaos, we
will show that the defect profiles, which constitute the phase twist
initial condition for the resulting daughter holes, show rather little
scatter for fixed $c_1$ and $c_3$. Changes in $c_1$ and $c_3$,
however, are encoded in changes in the defect profiles, and thus lead
to changes in the typical lifetimes of the daughter holes. We then
demonstrate that the chaos does not result from variations in defect
profiles. It rather follows from the sensitivity of the holes to the
states they invade, since the passage of each hole disorders the
background wavenumber profile leading to disordered background states.
{\it It is the self--disordering action of the holes that is primarily
responsible for the spatiotemporal chaos}.

With these insights, we can then construct a simplified lattice model
for hole--defect chaos, which both reproduces the correct qualitative
behavior as $c_1$ and $c_3$ are varied and which captures the correct
mechanism (propagating, self--disordering holes).  Our initial
findings on this subject can be found in Ref.~\cite{mvhmh}, where we
introduced the concept of self--disordering, and outlined a simplified
lattice model. However in this paper, we investigate the subject in
considerably greater depth, and, in particular, provide much more
conclusive evidence for the correctnesss of the self--disordering
hypothesis.

We now briefly summarize the structure of the paper. 
Topics discussed already in earlier work \cite{mvh,mvhmh} are
dealt with rather briefly. We start in Section~II by describing
hole--defect dynamics on a local scale. In Section~III, we then use
this knowledge to investigate disordered hole--defect dynamics, and
conclusively show that it is the coupling of the holes to a
self--disordered background which is the dominant mechanism for
spatiotemporal chaos. This concept is then illustrated by a minimal
lattice model for hole--defect dynamics in Section~IV, before we draw
our conclusions in Section~V.

\section{Hole--defect dynamics}

We begin by studying the hole lifetime $\tau$ as a function of the
initial conditions (Fig.~\ref{fig4}). This study motivates the central
question of this paper: how does $\tau$ depend on the initial
conditions and on the external wavenumber, and which of these
dependencies is most important for spatiotemporal chaos?  We then
study general properties of defect profiles, and demonstrate that in
hole--defect chaos the profiles of defects show rather little
scatter. We also show how the lifetimes of ``daughter'' holes born
from a typical defect vary with $c_1$ and $c_3$. Taken together, the
data presented here forms direct evidence for the heuristic picture of
hole--defect dynamics developed in Refs.~\cite{mvh,mvhmh}.

\subsection{Incoherent homoclons}\label{ssih}

In full dynamic states of the CGLE, one does not observe the unstable
{\em coherent} homoclinic holes unless one fine--tunes the initial
conditions (see below). Instead evolving {\em incoherent} holes which
either decay or start to grow towards defects occur \cite{mvh,mvhmh}.
Let us consider the short--time evolution of an isolated incoherent
hole propagating into a regular plane wave state. Holes can be seeded
from initial conditions like \cite{mvhmh}:
\begin{equation}
A=\exp(i[q_{ex}x+(\pi/2)\tanh(\gamma x)])~.\label{peakeq}
\end{equation}
The two essential parameters $\gamma$ and $q_{ex}$ represent,
respectively, the initial conditions from which the incoherent hole is
born and the background wavenumber of the state into which the hole
propagates. In this context, a single parameter $\gamma$ is sufficient
to scan through different initial conditions, since the coherent holes
have just one unstable mode \cite{mvh}.

\begin{figure}
\includegraphics[width=8.5cm]{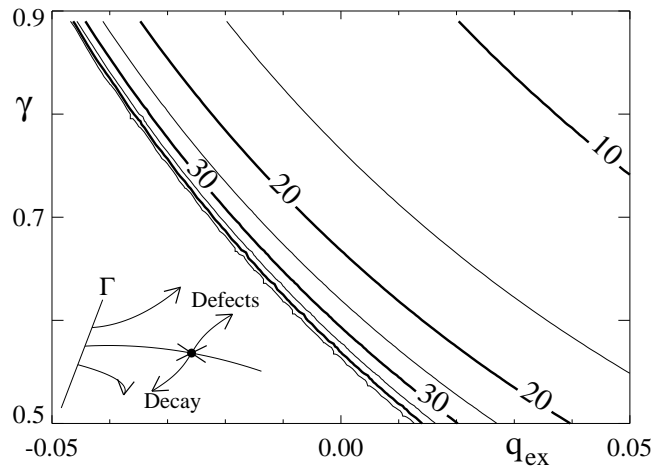}
\caption{Contour plot showing the lifetime $\tau$ of an initial
incoherent homoclon before a defect is generated. The initial
condition is given by Eq.(\ref{peakeq}), and the lifetime is plotted
as a function of $\gamma$ and $q_{ex}$. Note that the lifetime
diverges as $q_{ex}$ or $\gamma$ are reduced. In the left--bottom
corner of the diagram, the incoherent homoclon decays and no defects
are formed. The inset shows a sketch of the phase space around the
homoclon saddle (after Ref.~\cite{mvhmh}), where the manifold $\Gamma$
represents the family of peaked initial conditions of the form
(\ref{peakeq}).}
\label{fig4}
\end{figure}

A detailed contour plot of the lifetime of an initial incoherent hole
as a function of $\gamma$ and $q_{ex}$ is shown in
Fig.~\ref{fig4}. These results were obtained using a semi--implicit
numerical integration of the CGLE, with space and time increments
$dx=0.25$ and $dt=0.01$. As expected, three possibilities can arise
for the time evolution of the initial peak: evolution towards a defect
(upper right part of Fig.~\ref{fig4}), decay (lower left part of
Fig.~\ref{fig4}), or evolution arbitrary close to a coherent
homoclinic hole (the boundary between these two regions).  These
possibilities, together with an illustrative sketch of the phase
space, are shown in the inset of Fig.~\ref{fig4}. The rather simple
and monotonic behavior of $\tau$ with $q_{ex}$ and $\gamma$ is
somewhat of a surprise, and this reinforces our simple phase space
picture; no other solutions seem to be relevant in this region of
phase space.

Since homoclons are neither sinks nor sources, Fig.~\ref{fig4} can be
interpreted as follows: for a right--moving homoclon an incoming wave
with positive wavenumber tends to push the homoclon more quickly
towards a defect; previously we have referred to this as ``winding
up'' of the homoclon. Similarly, an incoming wave with negative
wavenumber ``winds down'' a right--moving homoclon, possibly even
preventing the formation of a defect \cite{mvh}.

\subsection{Defects}\label{def}

\begin{figure}
  \includegraphics[width=8.5cm]{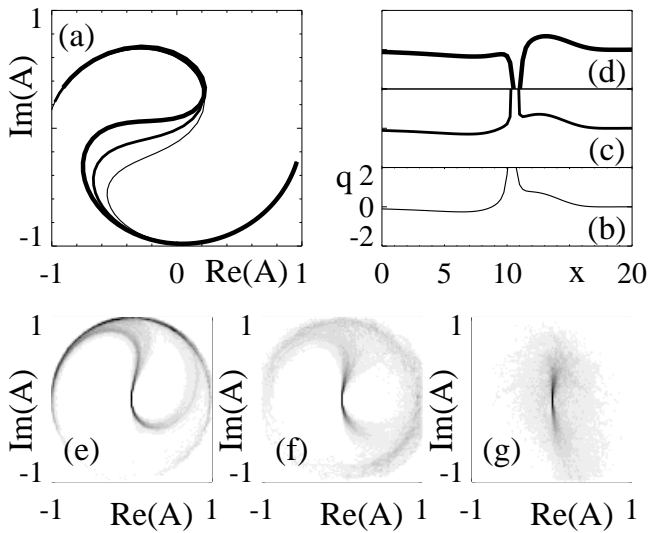} \caption{(a) Plots of the real
and imaginary part of $A$ around a defect, just before (thin line),
close to (medium line) and after (thick line) the defect has occurred;
the time difference between successive profiles is $0.01$.  Note that
in this representation the dynamics looks completely smooth. (b-d)
$q$-profiles for these three cases (identical vertical scale in each
case); just before the defect occurs, a large positive spike develops
in the profile, while after the defect this becomes a large negative
spike. (e-g) Statistics of defects obtained by overlaying $10^3$
defect profiles of spatial extension (width) 20, centered around the
defect position.
An arbitrary phase factor has been divided out by requiring that $Re(
\partial_x (A)|_{\rm{defect}})\!=\!0$. All data was collected in a
system of size 500, after a transient of 500. The coefficients $c_1$
and $c_3$ are: (e) $0.6,1.4$, (f) $1.4, 1.4$, (g) $3.0,3.0$. }
\label{fig34}
\end{figure}

We now study the defect profiles themselves in more detail.  In
Fig.~\ref{fig34}a we show complex plane plots of $Re(a)$ vs. $Im(A)$
just before, close to, and just after a defect. As can be seen, there
is no singular behavior whatsoever: the real and imaginary parts are
smooth functions of $x$ and $t$, even at the time of defect
formation. However, when transforming to polar coordinates, a
singularity manifests itself at the defect, where $|A|\to 0$. This can
also be seen from the $q$-profiles shown in Fig.~\ref{fig34}b-d. In fact,
it is straightforward to show that the maximum value of the local phase
gradient $q_{m}$ diverges as $(\Delta t)^{-1}$ at a defect \cite{mvhmh},
where $\Delta t$ is the time before defect nucleation (see Fig.~3b of
Ref.~\cite{mvhmh}) \cite{error}.

In Fig.~\ref{fig34}(e-g), we overlay complex plane plots of $A$ around
$10^3$ defects obtained from numerical simulations of the CGLE in the
chaotic regime.  Surprisingly (see Fig.~\ref{fig34}e), defect profiles 
of the interior chaotic states are dominated by a single profile in the 
hole--defect regime, similar to fixing $\gamma$ in Fig.~\ref{fig4}. 
This provides a strong indication that
hole--defect chaos does not come from scatter in the defect profiles.
For large enough $c_1$ and $c_3$, where holes no longer play a role,
and where hard defect chaos sets in \cite{maw}, the profiles show a
much larger scatter (Fig.~\ref{fig34}(f,g)).

\subsection{Defect $\to$ holes} \label{defhole}

Suppose a hole has evolved to a defect; what dynamics occurs after
this defect has formed? As Fig.~\ref{fig34}(b-d) shows, defects
generate a negative and positive phase--gradient peak in close
proximity. The negative (positive) phase gradient peak generates a
left (right) moving hole, and analogous to what we described in
Fig.~\ref{fig4}, the lifetimes of these holes depend on the initial
peak and on $q_{ex}$. Hence the defect profile acts as an initial
condition for its daughter holes, as can also clearly be seen in
Fig.~\ref{fig2}.

\begin{figure}  
   \includegraphics[width=8.5cm]{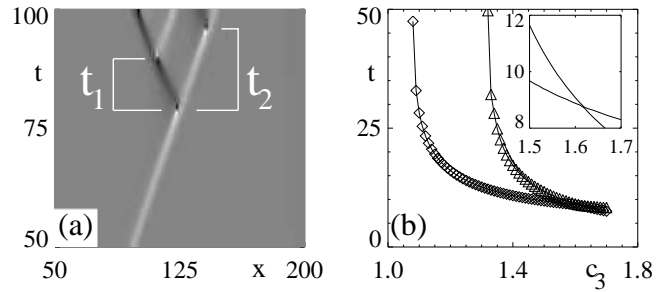} \caption{(a) Example of the
divergence of a near--coherent right--moving hole, showing also the
definition of times $t_1$ and $t_2$. (b) Times $t_1$ (diamonds) and
$t_2$ (triangles) as a function of $c_3$ for $c_1=0.6$. The inset
shows that the curves for $t_1$ and $t_2$ actually cross for
sufficiently large $c_3$. }\label{fig7}
\end{figure}

We now examine the fate of these daughter holes in the well--defined
case where the initial defect is generated from the divergence of a
right--moving, near--coherent homoclon in a $q_{ex}=0$ background
state (see Fig.~\ref{fig7}a). We then define $t_1$ and $t_2$ as the
lifetimes of the resulting daughter holes. When a daughter hole does
not grow out to form a defect, its lifetime diverges. In
Fig.~\ref{fig7}b we plot $t_1$ and $t_2$ for $c_1=0.6$ as a function
of $c_3$. The initial hole that formed the first defect has a lifetime
of at least $60$, and we have checked that a further increase of this
time does not change $t_1$ and $t_2$ appreciably. When both $t_1$ and
$t_2$ are infinite, no defect sustaining states can be formed, and the
final state of the CGLE is in general a simple plane wave. When only
$t_1$ is finite, isolated zigzag states are formed; such states have
been discussed in Ref.~\cite{mads}, and we will see some examples
below.  When both $t_1$ and $t_2$ are finite, and of comparable value,
more disordered states occur. We will later use this data on $t_1$ and
$t_2$ to calibrate our minimal lattice model for spatiotemporal chaos.

Hence, we see that changes in $c_1$ and $c_3$ not only lead to changes
in the defect profiles, but also modify the lifetimes of the resulting
daughter holes. However, for fixed $c_1$ and $c_3$, we have seen that
the defect profiles, which act as initial conditions for the
daughter holes, show rather little scatter, at least in the
hole--defect regime. In the next section, we will build on this
knowledge to unravel the causes of hole--defect spatiotemporal chaos.
 
\section{Mechanism of hole--defect chaos}
\label{global}

In this paper and in earlier work \cite{mvhmh}, we have argued that
the principal cause of the spatiotemporally chaotic behavior in the
1D CGLE is the movement of holes through a self--disordered
background. Clearly, as we can see from Fig.~\ref{fig4}, a disordered 
background wavenumber $q_{ex}$ will give rise to varying hole lifetimes 
and thus to disordered hole--defect 
dynamics. In this section, we explicitly demonstrate the correctness of 
this mechanism by modifying the CGLE dynamics in two ways:

\noindent
{\em Model (i): Fixed defect profile}. Whenever a defect occurs, this
defect is replaced by a standardized defect profile (obtained from an
edge defect). Here the dynamics will be chaotic, showing the
irrelevance of the scatter in defect profiles.

\noindent
{\em Model (ii): Background between holes $\rightarrow$ plane wave
with $q\!=\!0$}. At each timestep, the background between any two
holes is replaced by a plane wave with wavenumber zero. Here no chaos
will occur, illustrating the crucial importance of the self--disordered
background.

\begin{figure} \vspace{0cm}
 \includegraphics[width=8.5cm]{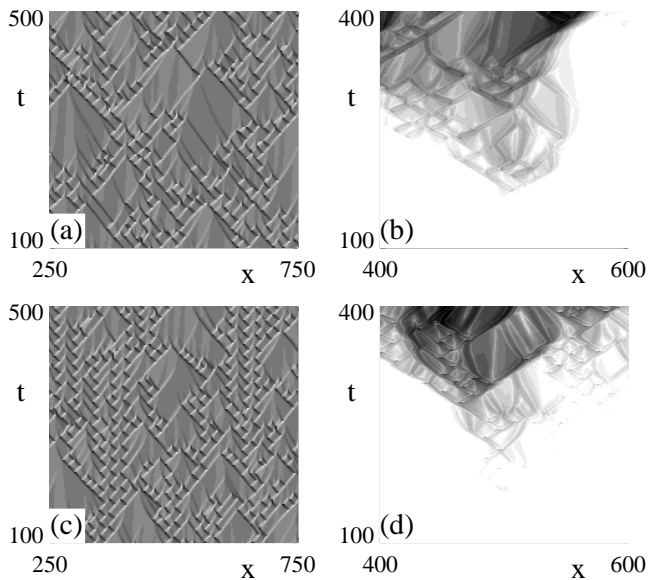}
\caption{(a) Space--time plot of $q$ in the ordinary CGLE for $c_1=0.6,
c_3=1.4$. (b) Log--greyscale plot of growth of perturbations. At
$t=166.66, x=500$, one gridpoint was altered by $10^{-6}$.  (c)
Space--time plot of $q$ in model {\em{(i)}}, the CGLE with fixed
defect profiles. (d) Log--greyscale plot of growth of perturbations
for the fixed defect model {\em{(i)}}.}
\label{figA}
\end{figure}

\begin{figure} \vspace{0cm}
 \includegraphics[width=8.5cm]{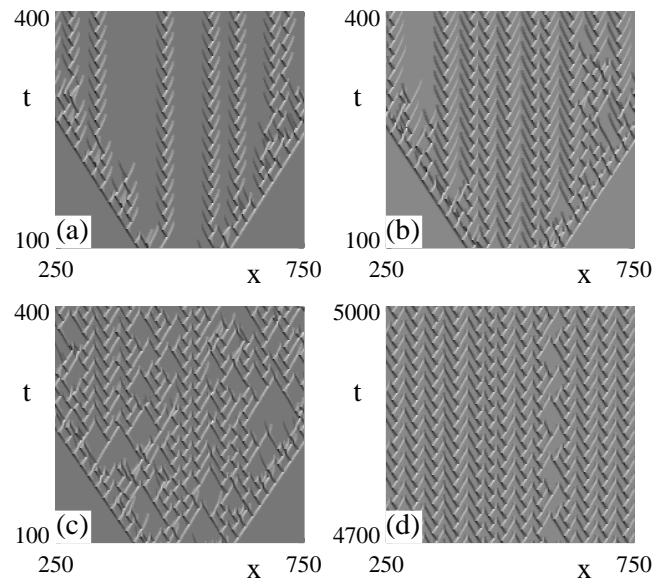}
\caption{Dynamical states in model {\em{(ii)}}, i.e., the CGLE where
the background between holes is replaced by a $q=0$ plane wave. (a)
For $c_3=1.4$ and $c_1=0.6$ only a few isolated zigzags occur. (b)
When $c_1$ is increased to $0.7$, more zigzags occur, but there is no
chaos. (c--d) For $c_1 = 0.8$, a disordered transient occurs (c) that
eventually freezes into a quasiperiodic zigzag state without disorder
(d).}
\label{figB}
\end{figure}

In case {\em{(i)}}, the size of the replaced defect profile was five
centered around the defect; in case {\em{(ii)}} the background was
defined to be all regions where $|A|>0.95$.
Our results are substantially independent of the exact defect size or
cutoff value. In both cases, it is crucial to ensure that no jumps in
the phase occur at the edges of the replaced regions. This can be
achieved by phase matching the replaced region (either defect profile
or plane wave) at the left boundary, while the state to the right of
the replaced region is multiplied by a phase factor to enforce
phase--continuity at the right edge. We take open boundary conditions
(i.e. $\partial_x A=0$) and only study the behavior far away from
these boundaries.

In Fig.~\ref{figA} we show an example of the dynamics and spreading of
a localized perturbation for the full CGLE (Fig.~\ref{figA}a--b) and
for the ``fixed defect'' model {\em{(i)}} (Fig.~\ref{figA}c--d), both
for $c_1\!=\!0.6, c_3\!=\!1.4$. For both models, we took as an initial
condition a defect rich state, which after a few timesteps shows the
typical hole--defect dynamics. At $t\!\approx\!167$ we applied a local
perturbation of strength $10^{-6}$ to the middle gridpoint
(corresponding to $x\!=\!500$) and followed the evolution of both the
perturbed and the unperturbed systems in order to follow the
spreading of perturbations. For the full CGLE (Fig.~\ref{figA}a--b),
the perturbation spreads along with the propagation of the holes.  We
note that the initial growth of the perturbation manifests itself in
slight ``shifts'' of the spatial and temporal positions of the
defects. In particular when two holes collide, a strong amplification
of the perturbations is observed.

We can now compare this with the above fixed defect profile model {\em
(i)} (Fig.~\ref{figA}c,d). Clearly, the replacement of the defects
does not destroy the chaotic behavior of the system, as confirmed by
the spreading of a localized perturbation (Fig.~\ref{figA}d), which
propagates in a similar fashion to the full CGLE
(Fig.~\ref{figA}b). This strongly indicates that variation in the
defect profiles is not contributing in a major way to the
spatiotemporally chaotic behavior of the full CGLE. We should also
point out one subtlety here: due to the discretization of time, the
times at which defect profiles are replaced are also discretized, and
one may worry whether this destroys the chaotic properties of the
model. However, we have performed simulations for a smaller timestep
($dt=0.001$) and found no qualitative difference. As we will see, this
issue of discretization will play a more important role in the lattice
model discussed in section \ref{lattice}.

Turning now to model (ii), where laminar regions of the CGLE are
replaced by $q=0$ plane waves, we see that the disorder is
destroyed. This is illustrated in Fig.~\ref{figB}, where we show
examples of model (ii) dynamics for $c_3\!=\!1.4$ and
$c_1\!=\!0.6,0.7,0.8$. Clearly chaos is suppressed for $c_1\!=\!0.6$
and 0.7 (Fig.~\ref{figB}a--b), with zigzag type patterns being
especially dominant. For $c_1\!=\!0.8$ (Fig.~\ref{figB}c), the initial
dynamics do appear disordered, but after a transient the system
evolves to a regular zigzag state (Fig.~\ref{figB}d).

From the behavior of models {\em{(i)}} and {\em (ii)} we conclude that
the self--disordered background is an essential ingredient for 
hole--defect chaos, while scatter in the
defect profiles is not.

\begin{figure} \vspace{.cm}
 \includegraphics[width=5.5cm]{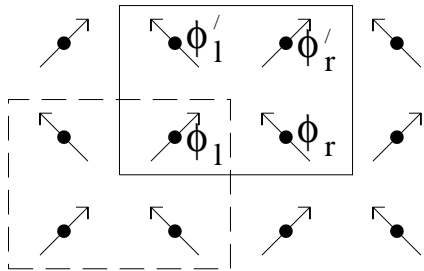}
\caption{Grid model geometry showing the sites (dots) and hole
propagation directions (arrows). The update rule is defined within a
$2\times 2$ cell, mapping $(\phi_l,\phi_r) \rightarrow
(\phi'_l,\phi'_r)$.  ``Active sites'' where $|\phi|>\phi_t$ represent
holes, while ``inactive sites'' where $|\phi|<\phi_t$ represent the
diffusive background.  When both sites are inactive the relevant
dynamics is phase diffusion: $\phi'_r \!=\!  D \phi_l \!+\!  (1\!-\!D)
\phi_r$. The value of $D$ is fixed at $0.05$ and is not
very important. When both sites are active, two holes collide and
merge: $\phi'_r \!=\!  \phi'_l \!=\!  (\phi_l \!+\!  \phi_r)/2$. When
one site is active but smaller than $\phi_d$, we implement the
evolution \cite{noteA}: $\phi'_r\!=\phi_l+\!
\lambda(\phi_l-\phi_n-g\phi_r)$ (we assume here that we have a right
moving hole, the case of a left moving hole follows by symmetry). Here
$\lambda$ sets the time scale and can be taken small (fixed at
$0.1$). When $\phi> \phi_d$ a defect occurs and two new holes, i.e.,
active sites, are generated; for details see text. }
\label{fig99}
\end{figure}

\section{Lattice model}\label{lattice}

To further justify and test our picture of self--disordered dynamics,
we will now combine the various hole--defect properties with the
left--right symmetry and local phase conservation to form a minimal
model of hole--defect dynamics. The model reproduces regular edge
states, spatiotemporal chaos and can be calibrated to give the correct
behavior as a function of $c_1$ and $c_3$. An earlier version of the
model was presented in Ref.~\cite{mvhmh}. However, as will become
clear, we have now modified and improved the model, and also for the
first time made direct comparisons with the full CGLE.

From our earlier analysis (see also Ref.~\cite{mvhmh}), we see that
the following hole--defect properties must be incorporated in the
model: {\em{(i)}} Incoherent holes propagate either left or right with
essentially constant velocity. {\em{(ii)}} Their lifetime depends on
$c_1$, $c_3$, and the wavenumber of the state into which they propagate.
When the local phase gradient extremum diverges, a defect
occurs. {\em{(iii)}} Each defect, in turn, acts as an initial
condition for a pair of incoherent holes.

In our lattice model we discretize both space and time, and take a
``staggered'' type of update rule which is completely specified by the
dynamics of a $2 \times 2$ cell (see Fig.~\ref{fig99}). We put a
single variable $\phi$ on each site, where $\phi$ corresponds to the
phase difference (the integral over the phase--gradient $q$) across a
cell, divided by $2 \pi$.  Local phase conservation is implemented by
$\phi'_l\!+\!\phi'_r \!=\!\phi_l\!+ \!\phi_r$, where the primed
(unprimed) variables refer to values after (before) an update.  Holes
are represented by active sites where $|\phi|>\phi_t$; here $\phi$
plays the role of the internal degree of freedom. Inactive sites are
those with $|\phi|<\phi_t$, and they represent the background. The
value of the cutoff $\phi_t$ is not very important as long as it is
much smaller than the value of $\phi$ for coherent holes, and is here
fixed at $0.15$. Without loss of generality we force holes with
positive (negative) $\phi$ to propagate only from $\phi_l$ ($\phi_r$)
to $\phi'_r$ ($\phi'_l$).

The details of the translation of these rules into the model can be
found in the caption of Fig.~\ref{fig99} and in Ref.~\cite{mvhmh},
with one exception. A ``defect'' is formed when $\phi_l \!>\!
\phi_d$. Here we have adopted two alternative schemes. In the simplest
case {\em{(A)}} (studied before in Ref.~\cite{mvhmh}), we take
$\phi'_r\!=\!\phi_{ad}$, and $\phi'_l\!=\!  \phi_d-1-\phi_{ad}$. Here
we completely fix the new holes. The factor $-1$ reflects the change
in the total winding number associated with the defects. Notice that
the overall winding number does not change by {\em exactly} $-1$.
This is because $\phi_l+\phi_r$ is usually slightly different from
$\phi_d$. As we will discuss below, to avoid breaking this ``phase
conservation'' we have also studied case {\em{(B)}}, where we take
$\phi'_r\!=\!\phi_{ad}$, and $\phi'_l\!=\!
\phi_l+\phi_r-1-\phi_{ad}$. Here some (small) scatter in the defect
profiles is allowed, but the change in the overall winding number is
now strictly $-1$.

The model does contain a large number of parameters, $g$, $\phi_n$,
$\phi_{d}$ and $\phi_{ad}$. We will first discuss the role of $g$ and
the difference between the two defect rules {\em{(A)}} and {\em{(B)}}.

\begin{figure} 
 \includegraphics[width=8.5cm]{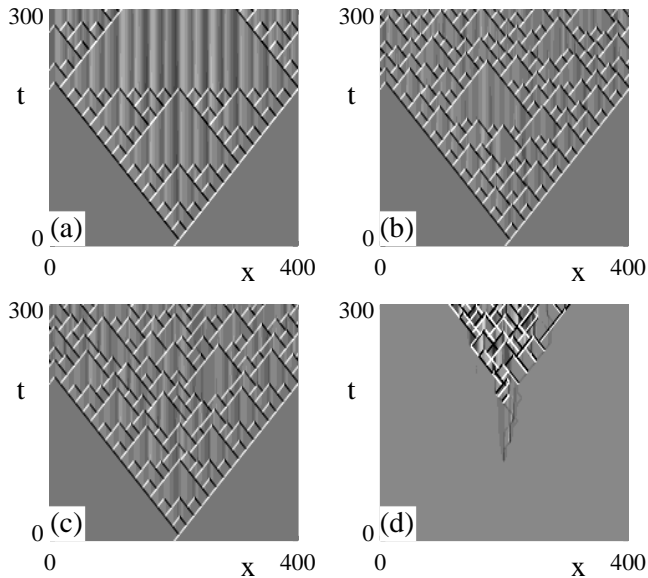}
\caption{Illustration of the necessary ingredients for disorder and
chaos in the grid model. 
In all cases, $\lambda=0.1, \phi_{ad}=0.75$,
$\phi_n=0.59$, $\phi_d=1.01$ (this leads to $t_1=10, t_2=12$, which is
the situation in the full CGLE for $c_1=0.6$ and $c_3=1.5$).  (a)
Without coupling to the background, $g=0$, the model with defect rule
{\em{(A)}} leads to regular Sierpinsky gaskets. (b) When $g=-3$, the
model generates disordered states, that are not strictly chaotic (see
text). (c) When defect rule {\em{(B)}} is applied, also for $g=-3$,
the dynamics is truly chaotic, as illustrated by the spreading of
perturbations (d).}
\label{fig15}
\end{figure}

In order for the model to reproduce the correct lifetime dependence of
edge--holes in hole--defect states, the coupling of the holes to their
background, $g$, should be taken negative (although its precise value
is unimportant).  For $g\!=\!0$ the hole lifetime $\tau$ becomes a
constant, independent of the $\phi$ of the state into which the holes
propagate, and moreover, the dynamical states are regular Sierpinsky
gaskets (Fig.~\ref{fig15}a).  For $g\!<\!0$ both the appropriate
$\tau$-divergence and disorder occur (Fig.~\ref{fig15}b), illustrating
the crucial importance of the coupling between the holes and the
self--disordered background.

It turns out that the model with defect rule {\em{(A)}} is not
strictly chaotic; sufficiently small perturbations do not always
chaotically spread. However, with defect rule {\em{(B)}} implemented,
infinitesimal perturbations do spread (see Fig.~\ref{fig15}c,d). To
understand this difference consider the fate of a tiny, localized
perturbation. Holes will sweep past and be influenced by this
perturbation, but since time is discrete, a sufficiently small
perturbation will not lead to a change in the time at which a hole
evolves to a defect. We have found that after a number of holes have
passed over such a perturbation, it can actually be absorbed, so that
no chaotic amplification occurs. It is therefore the combination of
the discreteness of time and the fixed defect profiles which do not,
strictly speaking, lead to chaos. By lowering $\lambda$, this problem
is diminished, but this makes the model far less effective
computationally. Alternatively, we have found that defect rule
{\em{(B)}} also circumvents this problem; perturbations can now never
be absorbed, due to the nature of the defect rule {\em{(B)}}.  In this
case the defect profile is not entirely fixed, but its scatter is
still rather small: for $\lambda=0.1$ a typical scatter is of the
order of 5\%, and this diminishes as $\lambda$ is decreased. Therefore
we can conclude that, in the continuous time limit of the lattice
model, the scatter of the defect profiles is not necessary to obtain
chaos. In the remaining part of the paper, we will use model
{\em{(B)}} only. 

\begin{figure} \vspace{.cm}
 \includegraphics[width=8.5cm]{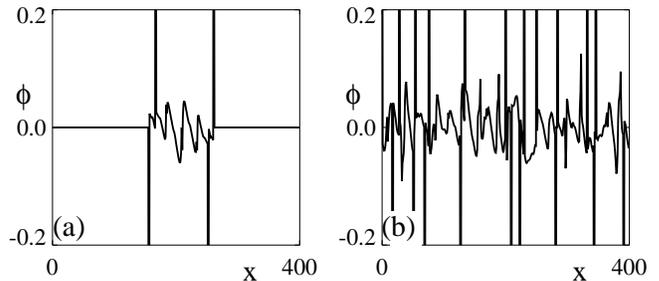}
\caption{Illustration of the self--disordering in two snapshots of the
field $\phi$ in our lattice model. Note that the scale is such that
only the background is clearly visible; the peaks correspond to
active, hole--like states. The data are taken from the runs shown in
Fig.~\ref{fig15}c, at (a) $t=60$ and (b) $t=300$.}
\label{FIGself}
\end{figure}

The self--disordering can be very clearly observed in the minimal
model, since its update rules unambiguously specify which sites are
``background'' and which are `active''. Two snapshots of the evolution
shown in Fig.~\ref{fig15}c, are plotted in Fig.~\ref{FIGself}. These
snapshots clearly demonstrate how, after sufficient time has passed,
the ``inactive'' background between the holes has become completely
disordered.

The essential parameters determining the qualitative nature of the
overall state are $\phi_n$, $\phi_d$ and $\phi_{ad}$.  These
parameters determine the amount of phase winding in the core of the
coherent holes with $q_{ex}\!=\!0$ ($\phi_n$), and in the new holes
generated by the defects ($\phi_d$, $\phi_{ad}$). When varying the
CGLE coefficients, these parameters change too, leading to
qualitatively different states.  In particular, they determine the
times $t_1$ and $t_2$ that we already studied for the full CGLE in
section \ref{defhole}.  We found that when $\phi_n$ and
$\phi_{ad}$ are both decreased, $t_1$ and $t_2$ roughly remain the
same. We have therefore kept $\phi_{ad}=0.65$, and varied $\phi_n$ and
$\phi_d$ to tune the values of $t_1$ and $t_2$. Notice that the
symmetry (or asymmetry) of the defect profile depends on $\phi_d-1$; a
value of $\phi_d <1$ typically promotes zigzag patterns. We have
determined the appropriate values of $\phi_n$ and $\phi_d$ for three
concrete cases, tabulated in Table~I. Notice that the parameters for
$c_3=1.5$ precisely correspond to those used in Fig.~\ref{fig15}.

\begin{table}[t]
\begin{ruledtabular}
\begin{tabular}{|c|c|c|c|c|c|}
  $c_1$ & $c_3$ & $t_1$ & $t_2$ &  $\phi_n$ & $\phi_d$ \\
  \hline
~0.6~ & ~1.25~ & ~14~ & ~$\infty$~ &~0.787~ & ~0.932~   \\
0.6 & 1.4 & 11 & 17 & 0.686 & 0.973  \\
0.6 & 1.5 & 10 & 12 & 0.59 & 1.01  
\end{tabular}
\end{ruledtabular}
 \caption{ Times $t_1$ and $t_2$ as obtained in the full CGLE,
and appropriate coefficients $\phi_n$ and $\phi_d$ that reproduce
these times in our grid model.}
\end{table}

\begin{figure} \vspace{.cm}
 \includegraphics[width=8.5cm]{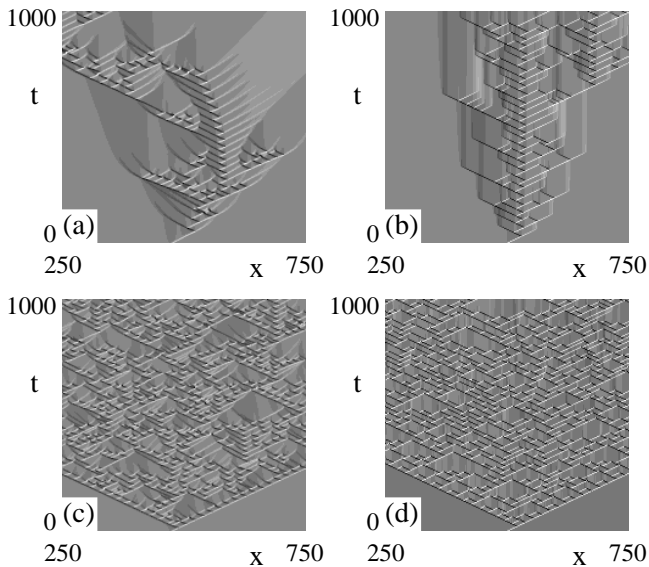}
\caption{Comparison between the dynamics of the full CGLE (a,c) and
our grid model (b,d), where $t_1$ and $t_2$ are matched according to
Table~I, with $c_1=0.6$, and (a-b) $c_3=1.25$ or (c-d) $c_3=1.4$.}
\label{FIG13}
\end{figure}

As can be seen in Fig.~\ref{FIG13}, the agreement between the simple
model and the CGLE is satisfactory, although clearly the CGLE displays
richer behavior.  Note that in the full CGLE, small perturbations of
the background wavenumber evolve in a non--trivial manner. For
example, a nonzero average background wavenumber introduces a drift of
the phase perturbations in the background between the holes
\cite{torcinidrift}. Since this phase dynamics is much slower than the
hole--defect dynamics, we have chosen to ignore it in the grid model,
and this accounts for the difference between Fig.~\ref{FIG13}a and
Fig.~\ref{FIG13}b.

Finally, we emphasize that the grid model allows us to disentangle the
mechanism of hole--defect chaos, by enabling us to completely control
the behavior of defects and the coupling between holes and the laminar
background. The grid model also has the advantage of being possibly
the simplest model that captures the essence of the self--disordered
hole-defect spatiotemporal chaos. We also emphasize that we have, for
the first time, carried out a detailed comparison between the full
CGLE and the grid model, both in our analysis of the spreading of
perturbations and in the calibration of the grid model as a function
of $c_1$ and $c_3$. Given the simplicity of the model, the agreement
with the full CGLE is striking.

\section{Conclusion}\label{discussion}

In conclusion, we have studied in depth the dynamics of local
structures in the 1D CGLE.  We have presented strong evidence that the
origin of the chaotic behavior in the 1D CGLE lies in the
self--disordering action of the holes, rather than in the scatter of
the defect profiles. Using this insight, we have then developed a
minimal lattice model for spatiotemporal chaos, which, despite its
simplicity, reproduces the essential spatiotemporally chaotic
phenomenology of the full CGLE.

How general are these results? We conjecture that there are two
crucial properties needed for hole--defect type chaos: propagating
saddle--like coherent structures (the holes) and a ``conserved'' field
(the phase field). Of course, the phase is not strictly conserved here
due to the occurrence of defects, but $(\int dx~ \psi) ~ \mbox{mod } 2
\pi $ is a conserved quantity. It is this conservation that is weakly
broken in our grid model for defect rule (A), but is preserved for
defect rule (B). Only the latter is truly chaotic. The conservation is
also the underlying reason why an evolving hole leaves an
inhomogeneous and self--disordered trail behind. Without such
a conserved field, there is no reason for ``self--disordering'' to
occur, and the holes then typically will exhibit a fixed lifetime,
leading to Sierpinsky gasket type patterns as is often the case in
reaction--diffusion models \cite{rd}. A related scenario appears to
occur in the periodically forced CGLE \cite{chatepikovsky}.  Conserved
fields of the type described here may be expected more generally for
systems undergoing a Hopf bifurcation, and can therefore be expected
to also occur in Ginzburg--Landau type equations including higher
order terms, and also in experiments. We have argued in
Ref.~\cite{grr}, that saddle--type structures like the homoclons here,
may be much more general. This leads us to believe that the type of
dynamics described here is not an artefact of the pure CGLE, but could
be far more widespread.

Our work opens up the possibility for quantitative studies of
hole--defect and homoclon dynamics, states which have recently been
observed in various convection experiments \cite{garnier,willem}. We
hope that our simple picture will advance these experimental studies
of space--time chaos into the quantitative realm. Local dynamics of
the type studied here, such as the dependence of lifetime on initial
conditions \cite{grr}, or measurements of quantities like the daughter
hole lifetimes $t_1$ and $t_2$ should be accessible in experiment,
thereby circumventing the difficulties normally associated with
characterizing fully--developed chaotic states.

Finally, we mention another commonly observed type of spatiotemporal
chaos occurring in systems when a periodic state undergoes a certain
symmetry breaking bifurcation \cite{faivre}.  Mathematically, such
systems may be described by a complex Ginzburg--Landau equation,
coupled to a phase field. Such models are sometimes referred to as
$A-\phi$ models \cite{aphi}.  Theoretically, the role of holes and
defects has not yet been studied in great detail for these systems,
but the main ingredients for hole--defect chaos of the type described
here appear to be present. We hope that our work will encourage further
studies in this area.

M.H. acknowledges support from Stichting FOM and from The Royal
Society.


\begin{references}

\bibitem{CH} M. C. Cross and P. C. Hohenberg, Rev. Mod. Phys. {\bf
65}, 851 (1993).

\bibitem{AK} I. S. Aranson and L. Kramer, Rev. Mod. Phys. {\bf 74}, 99
(2002).

\bibitem{chao1}  B. I. Shraiman,  A. Pumir,  W. van Saarloos,
P. C. Hohenberg,  H. Chat\'e and M. Holen, Physica D {\bf 57}, 241
(1992).

\bibitem{phasediagram} H. Chat\'e, Nonlinearity {\bf 7}, 185 (1994).

\bibitem{mvh} M. van Hecke, Phys. Rev. Lett. {\bf 80}, 1896 (1998).

\bibitem{maw} L. Brusch {\em et al.}, Phys. Rev. Lett. {\bf 85}, 86
(2000).

\bibitem{mvhmh} M. van Hecke and M. Howard, Phys. Rev. Lett. {\bf 86},
2018 (2001).

\bibitem{error} Note that the scale of the $\Delta t$ axis
in Fig.~3b of Ref.~\cite{mvhmh} should be a factor of
$10$ smaller than labeled.

\bibitem{mads} M. Ipsen and M. van Hecke, Physica D {\bf 160}, 103
(2001).

\bibitem{noteA} This equation is a simplified version of the quadratic
evolution equation for the holes introduced in Ref.~\cite{mvhmh}. This
quadratic equation describes the finite time divergence of the local
phase gradient extremum $q_m$ as a hole evolves towards a
defect. However, even though $q_m$ diverges at a defect, its local
integral does not.  Hence, the finite time divergence of the
local phase gradient maximum $q_m$ that signals a defect, can be
replaced by a cutoff $\phi_d$ for $\phi$ \cite{mvhmh}.

\bibitem{torcinidrift}
T. Kawahara, Phys. Rev. Lett. {\bf 51}, 381 (1983);
A. Torcini,  Phys. Rev. Lett. {\bf 77}, 1047 (1996).

\bibitem{rd} W. N. Reynolds {\em et al.}, Phys. Rev. Lett. {\bf 72},
2797 (1994); M. Zimmermann {\em et al.}, Physica D {\bf110}, 92
(1997); A. Doelman {\em et al.}, Nonlinearity {\bf 10}, 523 (1997);
Y. Hayase and T. Ohta, Phys. Rev. Lett. {\bf 81}, 1726 (1998); Y.
Nishiura and D. Ueyama, Physica D {\bf 130}, 73 (1999).

\bibitem{chatepikovsky} H. Chat\'e, A. Pikovsky and O. Rudzick,
  Physica D {\bf 131}, 17 (1999).

\bibitem{grr} M. van Hecke, Physica D {\bf 174}, 134 (2003).

\bibitem{garnier} N.  Garnier, A. Chiffaudel, F. Daviaud and
A. Prigent, Physica D {\bf 174}, 1 (2003); N. Garnier, A. Chiffaudel and
F. Daviaud, Physica D {\bf 174}, 30 (2003).

\bibitem{willem} L. Pastur, M. T. Westra and W. van de Water, Physica
D {\bf 174}, 71 (2003); L. Pastur, M. T. Westra, D. Snouck, W. van de
Water, M. van Hecke, C. Storm and W. van Saarloos, Phys. Rev. E {\bf 67},
036305 (2003).

\bibitem{faivre} S. Akamatsu and G. Faivre, Phys. Rev. E, {\bf 58},
3302 (1998); P. Brunet, J. M. Flesselles and L. Limat, Europhys. Lett.
{\bf 56}, 221 (2001).

\bibitem{aphi} P. Coullet and G. Iooss, Phys. Rev. Lett. {\bf 64} 866
(1990).

\end{references}
\end{document}